\documentclass[preprintnumbers,preprint,aps,amsmath,amssymb]{revtex4}
\usepackage{amsmath, amssymb, graphics,epsfig}
\newcommand{\mathsym}[1]{{}}

\def\lsim{\:\raisebox{-1.1ex}{$\stackrel{\textstyle<}{\sim}$}\:}
\def\gsim{\:\raisebox{-1.1ex}{$\stackrel{\textstyle>}{\sim}$}\:}
\def\10{$SO(10)$}
\def\21{SU(2) $\otimes$ U(1) }

\def\422{$SU(4) \otimes SU(2) \otimes SU(2)$}
\def\321{SU(3) $\otimes$ SU(2) $\otimes$ U(1)}
\def\lsim{\raise0.3ex\hbox{$\;<$\kern-0.75em\raise-1.1ex\hbox{$\sim\;$}}}
\def\gsim{\raise0.3ex\hbox{$\;>$\kern-0.75em\raise-1.1ex\hbox{$\sim\;$}}}
\def\vev#1{\left\langle #1\right\rangle}%

\newcommand{\ba}{\begin{array}}
\newcommand{\ea}{\end{array}}
\newcommand{\be}{\begin{equation}}
\newcommand{\ee}{\end{equation}}
\newcommand{\beqa}{\begin{eqnarray}}
\newcommand{\eeqa}{\end{eqnarray}}
\relax
\def\321{$SU(3)\times SU(2)\times U(1)$}

\newcommand{\dms}  {\Delta m^2_{sol}}
\newcommand{\Dma}  {\Delta m^2_{atm}}

\begin{document}
\bigskip
\title[]{Quasi-degenerate neutrinos in $SO(10)$}
\author{
Anjan S. Joshipura\footnote{anjan@prl.res.in} and Ketan M.
Patel\footnote{kmpatel@prl.res.in}}
\affiliation{
Physical Research Laboratory, Navarangpura, Ahmedabad 380 009, India
\vskip 2.0truecm}
\begin{abstract}
\vskip 1.0 truecm
Quark lepton universality inherent in grand unified theories based on $SO(10)$ gauge group generically leads to
hierarchical neutrino masses. We propose a specific ansatz for the structure of Yukawa matrices in $SO(10)$ models which 
differ from this generic expectations and lead to quasi degenerate neutrinos through the type-I seesaw mechanism.
Consistency of this ansatz is demonstrated through a detailed fits to fermion masses and mixing angles all of which can be 
explained with reasonable accuracy in a model which uses the Higgs fields transforming as $10,120$ and $\overline{126}$ representations
of $SO(10)$. The proposed ansatz is shown to follow from an extended model based on the three generations of the vector like fermions 
and an $O(3)$ flavour symmetry. Successful numerical fits are also discussed in earlier proposed models which used combination of the type-I and type-II seesaw mechanisms for obtaining quasi degenerate neutrinos. Large neutrino mixing angles emerge as a consequence of neutrino mass degeneracy
in both these cases.

\end{abstract}

\maketitle
\section{Introduction}
Experiments over the years have revealed that \\
(1) Two of the neutrino mixing angles are large as opposed to the small quark mixing angles. \\
(2) Neutrino mass hierarchy is milder compared to quarks and extreme  case of all neutrinos being quasi degenerate is still an allowed possibility.\\
Several independent reasons have been advanced\cite{rev,btau,as} to understand feature (1) of the fermion spectrum but it  may be that its answer lies  in (2).  Large mixing angles become quite natural if neutrinos are almost degenerate. They remain undefined in the exact degenerate limit.  A small  perturbation which leads to differences in neutrino  masses can also stabilize all or some of the mixing angles to large values. Thus theory which predicts quasi degeneracy has built in mechanism to explain large mixing angles. We present an \10 based unified description of fermion masses and mixing leading to hierarchical charged fermions and quasi degenerate neutrino masses. Large neutrino mixing emerges as a consequence of neutrino mass degeneracy.

\10 models provide a natural framework for understanding neutrino masses because of the  seesaw mechanisms \cite{rev}  inherent in them. Neutrino masses arise in these models from two separate sources either from the vacuum expectation value of the left-handed triplet (type-II) or from the right handed triplet (type-I) Higgs. It was pointed out\cite{caldmoh,valle} long ago that combination of these two sources provides an interesting framework for understanding quasi degeneracy of neutrinos. In this approach, some flavour symmetry leads to degenerate type-II contribution and its breaking in the Dirac neutrino masses then leads to departure from degeneracy through the type-I contribution. This is realizable if type-II contribution dominates over the type-I which is not always the case\cite{problem,fits}. Alternative possibility is that both degeneracy and its breaking arise from a single source namely type-I seesaw mechanism. This however requires a peculiar structure for the RH neutrino mass matrix $M_R$. It has been pointed out that the required  structure can arise  from the ``Dirac screening'' \cite{ds} or more generally from the application of the Minimal Flavour Violation \cite{mfv} hypothesis to the leptonic sector \cite{apv}.

While these possibilities are known there does not exist a detailed study of  all fermion masses and mixing in the context of realistic $SO(10)$ models with quasi degenerate neutrinos and we address this question using (A) type-I mechanism alone and (B) combination of  type-I and type-II mechanisms.

We use supersymmetric $SO(10)$ as our basic framework. Fermion masses arise in renormalizable $SO(10)$ models through their couplings to Higgs fields transforming as $10$,$\overline{126}$ and $120$ representations. One needs at least two of these fields to get fermion mixing and the minimal model with $10$ and $\overline{126}$ has attracted a lot of attention\cite{btau,problem,fits,minimal}. There have been studies of models with additional $120$ also\cite{nonminimal,grimus,akp}. In our context, we find that all the three Higgs representations are needed to obtain satisfactory fits to  fermion masses and mixing. Starting with a supersymmetric $SO(10)$, an effective Minimal Supersymmetric Standard Model (MSSM) is obtained by assuming fine tuning which keeps only two Higgs doublets lights. The final fermion mass matrices obtained after $SO(10) $ and $SU(2)_L\times U(1)$ breaking can be parametrized as   \cite{grimus,akp}:
\beqa \label{matrices}
M_d&=& H+F+ G~,\nonumber \\
M_u&=&r (H+s~ F+ t_u~ G), \nonumber\\
M_l&=& H-3~ F+~t_l~ G,\nonumber\\
M_D&=&r (H-3s~ F+t_D~ G),\nonumber\\
M_L&=& r_L ~F,\nonumber\\
M_R&=& r_R^{-1}~ F,\eeqa
where the matrices $H$ and $F$ are complex symmetric and $G$ is an anti-symmetric matrix in generation space. $H,F,G$ arise from the fermionic Yukawa couplings to the $10,\overline{126},120$ Higgs fields respectively. $r,~s,~t_u,~t_l,~t_D,~r_L,~r_R$ are complex parameters determined by the ratios of vacuum expectation values (vevs) and mixing among various Higgs doublets.
The light neutrino mass matrix is given by
\be \label{mnu}
{\cal M}_\nu=r_L F-r_R M_D F^{-1} M_D^T \equiv {\cal M}_\nu^{II}+{\cal M}_{\nu}^{I}~. \ee

It is known that the above fermion mass structure allows different mixing patterns for quarks and neutrinos
if type-II seesaw mechanism dominates \cite{btau,minimal}. Consider the limit in which the contribution of the 10-plet $H$ dominates.
In this limit, all the charged fermions are diagonalized by the same matrix and the Cabibbo-Kobayashi-Maskawa (CKM) matrix
becomes proportional to identity. In the same limit, neutrino mixing with the type-II dominance is governed by $F$ in eq.(\ref{mnu})
leading to non-trivial leptonic mixing. In fact, if only $H$ dominates the charged fermion masses then one can obtain $b$-$\tau$
unification which in turn drives the large atmospheric mixing \cite{btau}. The existing fits \cite{fits,akp} to fermion masses and mixing with type-II dominance  are for the hierarchical neutrino masses. Degenerate neutrino spectrum can be obtained in this approach with
an additional assumption:
\be \label{type2deg}
F=c_0 I \ee
$I$ denoting a $3\times 3$ identity matrix. The sub-dominant type-I contribution can then lead to the quark mixing and neutrino mass differences.

The realization of the attractive type-II dominated scenario was found difficult in the context of the minimal model \cite{problem,fits}. It was found that parameter space  favored by the overall fit to fermion masses  suppresses the type-II contribution compared to the type-I. This motivates us to study
degenerate neutrinos in the context of a purely type-I seesaw mechanism. A general framework based on flavour symmetry arguments
to obtain quasi degenerate neutrinos in type-I seesaw was recently discussed \cite{apv}. It was found  that the right handed neutrino mass matrix in an effective theory invariant under a spontaneously broken flavour symmetry  $G_f=O(3)_l\times O(3)_e\times O(3)_\nu\times U(1)_R$ has a dominant term
\be M_R\approx M_D^T M_D+... ~,\ee
where each $O(3)_f$ acts on the flavour spaces of the lepton doublet ($f=l$), the right-handed electrons ($f=e$) and the right handed neutrino ($f=\nu$) fields respectively and $U(1)_R$ corresponds to the right handed lepton number. The above form for $M_R$ leads to the degenerate neutrinos through  the type-I seesaw mechanism. An equivalent description at the $SO(10)$ level can be obtained  by imposing  the following ansatz:
\be \label{type1deg}
F=a H^2~,\ee
In the following, we will work out the detailed consequences of this ansatz for fermion masses and mixing and show that such an ansatz can be realized using a flavour symmetry. Since $H$ is a symmetric matrix it can be diagonalized by a unitary matrix.
\be\label {htra}
U^T~H~U=D_H~,
\ee
where $D_H$ is a diagonal matrix with real elements. Without loss of generality, we can express the mass matrices in (\ref{matrices}) in an \10  basis with a diagonal $H$. This basis are obtained from eq.(\ref{matrices}) by the replacement $H\rightarrow D_H$ and 
\be\label {ftra}
F\longrightarrow U^T~F~U = a (U^THU~U^{\dagger} U^{*} ~U^THU) = a ~ D_H V^{*}  D_H~.
\ee
$G$ retains its antisymmetric form and we use the same notation for it and for various mass matrices in the new basis. From now on, we will work in this rotated basis.
$V=U^TU$ in eq.(\ref{ftra}) is a symmetric unitary matrix which can be parametrized \cite{branco}  as
\be \label {param}
V=P R_{23}^T(\phi)U_{12}(\theta,\alpha)R_{23}(\phi)P ~,\ee
with
\be \ba{cc}
R_{23}(\phi)=\left( \ba{ccc}
1&0&0\\
0&\cos\phi&\sin\phi\\
0&\sin\phi&-\cos\phi \\  \ea \right)
&;U_{12}(\theta,\alpha)=\left( \ba{ccc}
\cos\theta&\sin\theta&0\\
\sin\theta&-\cos\theta&0\\
0&0&e^{i\alpha} \\  \ea \right)
\ea ~\ee
and $P=Diag(e^{i \alpha_1},e^{i \alpha_2},1)$ is a diagonal phase matrix (One phase in $P$ is absorbed in the complex parameter $a$ in eq.({\ref{ftra}})).

Before we present the detailed fits let us look at the implications of the ansatz eq.(\ref{type1deg}) qualitatively.
\begin{itemize}
\item Correct $b$-$\tau$ unification and second generation masses are obtained if dominant contribution to the charged fermion masses come from the 10-plet, i.e. from $H$ with a sub-dominant contribution from $\overline{126},120$ fields. Retaining only the $H$  contribution, the ansatz, eq.(\ref{type1deg}) implies that
\be \label{deg0}
  {\cal M}_{\nu}^{I}=- r_R M_D F^{-1} M_D^T\approx- \frac{r^2 r_R}{a}  V+...~,\ee
where the $...$ terms arise from the $\overline{126}$ and $120$ contribution to the Dirac mass matrix $M_D$. CKM matrix is unity in this limit while the neutrino mixing is determined from $V$. Diagonalization of $V$ leads \cite{branco} to $\theta_{23}=\phi$,
$\theta_{12}=\frac{\theta}{2}$ and $\theta_{13}=0$ where the 
angles $\theta_{ij}$ are angles defined in the standard parametrization of the leptonic mixing matrix in which $\theta_{12}$ drives the solar and $\theta_{23}$ the atmospheric neutrino oscillations. Thus ansatz in eq.(\ref{type1deg}) can lead to correct description of the quark and leptonic mixing angles to zeroth order without requiring the type-II dominance as is commonly done.
\item If $H$ in the original basis was real then $V$ entering eq.(\ref{ftra}) would be unity. In this case, all the fermion mixing vanish in the absence of 
the $120$ contribution. Thus complex couplings and CP violation proves to be important in understanding large neutrino mixing within this approach.
Numerically, we find that even after including $120$ contribution, one cannot get the correct mixing pattern with a real $H$. 
\end{itemize}
The mixing angles obtained at zeroth order with $H$ dominance get corrected by the contributions from $\overline{126}$ and $120$-plets. They induce non-zero quark mixing angles and perturb eq.(\ref{deg0}):
\beqa
\label{nu1X}
{\cal M}_\nu^I(M_X)&=&-\frac{r_R r^2}{a} (V - 6saD_H+t_D(GD_H^{-1}V-VD_H^{-1}G))+ {\cal O}(s^2, t_D^2))~,
\eeqa
The above neutrino mass matrix corresponds to an effective dimension five operator induced 
after integration of the right handed neutrino fields. Assuming that the heavy mass scale is close to the GUT scale and neglecting the effect of the 
Dirac neutrino couplings in the renormalization group (RG)  evolution one can  obtain \cite{rev}  the low scale neutrino mass matrix as follows.
Define the neutrino mass matrix ${\cal M}_{\nu f}(M_X)$ in the flavour basis as
\be \label{mnufx}
{\cal M}_{\nu f}(M_X) =U_l^\dagger{\cal M}_{\nu }(M_X) U_l^*~,\ee
where $U_l$ diagonalizes the charged lepton mass matrix $M_l$. The radiatively  corrected neutrino mass matrix is then given by \cite{rev}
 \be \label{mnufz}
{\cal M}_{\nu f}(M_Z) =I_\tau{\cal M}_{\nu }(M_X) I_\tau^{\dagger}~,\ee
where $I_\tau\approx Diag(1,1,1+\epsilon_\tau)$ and $ \epsilon_\tau\approx -\frac{1}{\text{cos}^2\beta} \frac{m_\tau^2}{16 \pi^2 \upsilon^2} \text{ln}\frac{M_X}{M_Z}$.
More detailed treatment then presented here would need to include  RH threshold and ruining between the GUT and the RH mass scale etc.

\section{Numerical Fits: Type-I seesaw}
We now discuss detailed fits to fermion masses and mixing based on the ansatz (\ref{type1deg}) and the fermion mass matrices, eq.(\ref{matrices}).
The latter are defined at the GUT scale $M_X$. We use as our input the quark and lepton masses obtained at $M_X$ in the MSSM for $\tan\beta=10$, $M_{SUSY}=1 \text{TeV}$ and  $M_{GUT}=2 \times 10^{16}\text{GeV}$ \cite{parida}. The quark mixing angles do not appreciably change compared to the low scale values and we use the specific values used in earlier analysis presented in \cite{grimus,akp}. Unlike in the previous works, the neutrino mixing angles are susceptible to change by the renormalization group (RG) evolution due to quasi degenerate nature of neutrinos. We include this effect as follows. Using the charged lepton mass matrix at $M_X$, we numerically determine the neutrino mass matrix in the flavour basis at $M_X$ through eq.(\ref{mnufx}).
Neglecting running of the mixing angles in $U_l$, the low scale neutrino mass matrix in  eq.(\ref{mnufz}) is numerically determined and  used to obtain the observable neutrino masses and mixing angles. For neutrino masses and lepton mixings, we use the updated low energy values given in \cite{numas}. All our input data (central values and $1 \sigma$ errors) are summarized in Table(\ref{tab:input}). 

\begin{center}
\begin{small}
\begin{table} [ht]
\begin{math}
\begin{array}{|c|c||c|c|}
\hline
 m_d \text{[MeV]}&  1.03\pm 0.41&\Delta m^2_{sol}[\text{eV}^2]&(7.59\pm 0.20)\times10^{-5} \\ 
m_s\text{[MeV]}&19.6\pm 5.2&\Delta m^2_{atm}[\text{eV}^2]&\left(2.51\pm 0.12\right)\times 10^{-3}\\
m_b\text{[MeV]} & 1063.6_{-86.5}^{+141.4} & \sin  \theta _{12}^q & 0.2243\pm 0.0016 \\
 m_u\text{[MeV]} & 0.45\pm 0.15 & \sin  \theta _{23}^q & 0.0351\pm 0.0013 \\
 m_c\text{[MeV]}&210.3273_{-21.2264}^{+19.0036} & \sin  \theta _{13}^q & 0.0032\pm 0.0005 \\
 m_t\text{[MeV]}&82433.3_{-14768.6}^{+30267.6}& \sin ^2 \theta _{12}^l & 0.32\pm 0.016 \\
 m_e \text{[MeV]}&0.3585\pm 0.0003 &\sin ^2 \theta _{23}^l&0.45\pm 0.092 \\
 m_{\mu }\text{[MeV]}&75.6715_{-0.0501}^{+0.0578} & \sin ^2\theta _{13}^l & < 0.049(3 \sigma) \\
m_{\tau }\text{[MeV]}&1292.2_{-1.2}^{+1.3}&\delta_{CKM}&60^{\circ }\pm 14^{\circ }\\
\hline
\end{array}
\end{math}
\caption{Input data used in this work (See text for details and references).}
\label{tab:input}
\end{table}
\end{small}
\end{center}

We do the $\chi^2$ fitting to check the viability of the model as previously done in \cite{fits,grimus,akp}. In this case we have total  25 real parameters (3 in $D_H$, 5 in $V$, 6 in $G$, real $r$, complex $s,~a,~t_u,~t_l,~t_D$ ) which are fitted over 16 observables (9 charged fermion masses, 4 CKM parameters, 2 leptonic mixing angles and $\dms/\Dma$ ).  Lepton mixings and $\dms/\Dma$  are independent of the overall neutrino mass ($m_0=|\frac{r_R r^2}{a}|$) appearing in eq.(\ref{nu1X}). This thus remains an arbitrary parameter and the overall degenerate mass scale cannot be fixed. We also set $r=\frac{m_t}{m_b}$ and minimize $\chi^2$ with respect to the remaining 24 parameters. The results of the minimization are displayed in Table(\ref{tab:2}). Three of the viable solutions are displayed. We obtained the best fit value of $\chi^2=2.038$ for which all the observables are fitted within $ \lesssim 0.9\sigma$. Solution(3) is also acceptable which fits all observables within $ \lesssim0.7 \sigma$ with exception of down quark mass $m_d$.
The fits obtained here are better than the one obtained by Bertolini {\it et al} \cite{fits} in case of the  minimal model with complete type-II seesaw dominance and hierarchical neutrinos. Unlike here, their fits have several observables which are $> 1$ or $2\sigma$ away from the central values. 

\begin{table} [ht]
\begin{small}
\begin{math}
\begin{array}{|c|c|c|c||c|c||c|c|}
\hline
 &    & \textbf{Sol. 1}  & \textbf{Sol. 1}  & \textbf{Sol. 2} & \textbf{Sol. 2}  & \textbf{Sol. 3} & \textbf{Sol. 3}   \\
 \text{No.} & \text{Observables} & \text{Fitted value} & \text{Pull} & \text{Fitted value} & \text{Pull} & \text{Fitted value} & \text{Pull}\\
\hline
 1 & m_d\text{[MeV]} & 0.653677 & -0.917861 & 0.678809 & -0.856564& 0.207819 &\textbf{ -2.00532} \\
 2 & m_s\text{[MeV]}  &17.5885 & -0.386821 & 22.8346 & 0.622041 & 21.6923 & 0.402361 \\
 3 & m_b\text{[GeV]}   & 1.11131 & 0.418721 &  0.94463 &\textbf{ -1.0440}& 1.05832 & -0.046348 \\
 4 & m_u\text{[MeV]} & 0.462718 & 0.0847896 & 0.461582 & 0.0772103 & 0.450825 & 0.00549932 \\
 5 & m_c\text{[GeV]}   & 0.210603 & 0.0136849 & 0.212603 & 0.113153 & 0.211727 & 0.0695654 \\
 6 & m_t\text{[GeV]}   & 63.6891 & -0.832404 & 56.7159 &\textbf{ -1.14208} & 67.6155 & -0.658038 \\
 7 & m_e\text{[MeV]}  & 0.358503 & 0.00969691 & 0.358516 & 0.0525082& 0.358506 & 0.0206782 \\
 8 & m_{\mu }\text{[MeV]}  & 75.6719 & 0.00734514 & 75.6765 & 0.0923818  & 75.6711 & -0.0083064 \\
 9 & m_{\tau }\text{[GeV]}  & 1.29219 & -0.00814429 & 1.2921 & -0.0804718 &1.29223 &0.0218404 \\
 10 & \frac{\dms}{\Dma} & 0.0303514 & 0.050109 & 0.0302197 & -0.00862837& 0.0303237 & 0.0377877 \\
 11 & \sin  \theta _{12}^q & 0.224205 & -0.0592102   & 0.224193 & -0.0666865& 0.224306 & 0.00359473 \\
 12 & \sin  \theta _{23}^q & 0.0351308 & 0.023704 & 0.0347491 & -0.269936 & 0.0350426 & -0.0441173 \\
 13 & \sin  \theta _{13}^q & 0.00319336 & -0.0132867  & 0.00322056 & 0.0411106 & 0.00315871 & -0.0825897 \\
 14 & \sin ^2 \theta _{12}^l & 0.319801 & -0.0619079 & 0.319568 & -0.076109& 0.321124 & 0.0187774 \\
 15 & \sin ^2 \theta _{23}^l & 0.481942 & 0.313909 & 0.437263 & -0.169787 & 0.436492 & -0.178126 \\
 16 & \sin ^2 \theta _{13}^l&\textbf{ 0.0195266} & -  & \textbf{0.0463404} &- &\textbf{ 0.00288176} & - \\
 17 & \delta _{CKM}[^{\circ}]  & 67.7227 & 0.247333 & 49.8678 & -0.422669 & 56.4935 & -0.134071 \\
 18 & \delta _{PMNS}[^{\circ}]   &\textbf{ 53.9786} & - &\textbf{ -52.7788} & - &\textbf{ -66.9939} & -\\
\hline
&\chi ^2   &   &2.038&   &3.844&   &4.684\\
\hline
\end{array}
\end{math}
\end{small}
\caption{Three best fit solutions for fermion masses and mixing obtained assuming the type-I seesaw dominance. Various observables
and their pulls obtained at the minimum are shown (See text for details). The bold faced quantities are predictions of the respective solutions.}
\label{tab:2}
\end{table}

We give below values of input parameters  determined from $\chi^2$ minimization in case of solution(1) of Table(\ref{tab:2}). 
\be
D_H={\rm Diag.} (0.00033658 , 0.0149966 , -0.757334) 
\ee 
\be
G_{12}=(0.00167+0.00154 I)~;~G_{13}=(0.0108+0.0101 I)~;~G_{23}=(0.191+0.033 I)~. \ee
\beqa
a&=&-(5.24 - 9.34 I) {\rm GeV}^{-1}; t_l=0.15 - 1.05 I; t_u=0.551- 0.084 I;  \nonumber \\
s&=&(-3.21 + 3.32 I)\times 10^{-4}; t_D=-(2.56 + 0.43 I)\times 10^{-4};   \nonumber \\
\phi&=&50.65^{\circ}; \theta=-56.9^{\circ}; \alpha=0.20^{\circ} ; \alpha_1=-39.55^{\circ};  \alpha_2=83.79^{\circ} 
\eeqa
Elements of $D_H$ and $G$ are expressed in GeV units.
Note that $\phi$ and $\theta/2$ respectively determine the atmospheric and the solar mixing angles at $M_X$ in the absence of perturbation.
These values get stabilized  at $M_X$ to $\phi\approx 40.4^\circ~;~\theta\approx -61.0^\circ$ once the perturbations from $\overline{126},120$ couplings are added. RG running changes them to the required values displayed in Table(\ref{tab:2}). $\theta_{13}$ has not been included in our definition of $\chi^2$ and its initial value was zero. This becomes non-zero but remains small in all the three solutions displayed. However,
almost the entire allowed range in $\theta_{13}$  is compatible with reasonable fits to other fermion masses as shown by the three solutions. Unlike the solar and atmospheric mixing angles, the ratio $\dms/\Dma$ changes appreciably from $0.09$ at $M_X$ to $0.03$ at $M_Z$ by the RG effects. All these solutions predict large CP violating leptonic phase. The heaviest RH neutrino mass scale following from the above numerical fit would be approximately (see eq.(\ref{matrices} and ansatz (\ref{type1deg}))
$$ M_3\approx r_R^{-1}|a| m_b^2\approx \frac{r^2}{m_0} m_b^2\approx 2 \times 10^{13} {\rm GeV}\left(\frac{0.2 {\rm eV}}{m_0}\right)~,$$
where  $m_0=\frac{r_R r^2}{|a|}$. Thus the RH neutrino mass falls below the GUT scale for this particular solution.

Let us now illustrate how the ansatz (\ref{type1deg}) can be obtained in  a model from a flavour symmetry.
A simple flavour symmetry to be used is $O(3)$ under which three generation of the 16-plet $\psi$ transform as triplets.
The $O(3)$ breaking is introduced through a complex flavon field $\eta$ transforming as spin 2.  We need to introduce three generations of vector-like multiplets $\Psi_V+\Psi_{\overline{V}}$
transforming as $(16,3)+(\overline{16},3) $ under $SO(10)\times O(3)$ and a $U(1)_X$ symmetry in order to realize eq.(\ref{type1deg}). The $X$-charges of 
$(\psi,\Psi_V,\Psi_{\overline{V}},\eta,\phi_{10},\phi_{\overline{126}})$ are chosen respectively as $(x,y,-y,1/2(y-x),-(x+y),-2 y)$ with $x\not= y$.
The general super potential invariant under $SO(10)\times O(3)\times U(1)_X$ can be written as:
\be \label{superp}
W=M\Psi_{\overline{V}}\Psi_V+\beta \Psi_V\Psi_V \phi_{\overline{126}}+\gamma \Psi_V\psi \phi_{10}+\frac{\delta}{M_P}\Psi_{\overline{V}}\eta^2\psi+
\frac{\delta'}{M_P}Tr\eta^2\Psi_{\overline{V}}\psi+.....\ee
The $O(3)$ and $U(1)_X$ breaking originates in the above super potential only from the Planck scale effects through the vev of the flavon field $\eta$. The last two terms are the only terms which determine both the 10 and $\overline{126}$ Yukawa couplings once the heavy vector like fields are integrated out.
The dotted terms correspond to terms suppressed by $M_P^2$.
Here, the mass $M$ of the vector like pair and the scale of the vev of $\eta$ lie above the GUT scale. The effective theory after integration of the vector like field is represented by
\be \label{Weffective}
W_{eff}\approx \frac{\beta\delta^2}{M^2M_P^2}\psi\xi^2\psi\phi_{\overline{126}}+\frac{\gamma\delta}{MM_P}\psi\xi\psi\phi_{10}~, \ee
where 
$$ \xi_{ab}\equiv \eta^2_{ab}+\frac{\delta'}{\delta}Tr\eta^2\delta_{ab}$$
and $a,b=1,2,3$ refer to the $O(3)$ index.
This effective super-potential is also $SO(10)\times O(3)\times U(1)_X$ invariant. The Yukawa coupling $H$ is proportional to the $\vev{\xi}$ and is  a general complex symmetric matrix. The $F$ is related to the square of $H$ and satisfies the ansatz in eq.(\ref{type1deg}). The coupling to the $120$ field can be generated by introducing a flavon field $\chi$ with the $U(1)_X$ charge-2 $x$ and transforming as a  triplet of $O(3)$. This leads to the Yukawa coupling matrix $G$ through the coupling
$$\psi \frac{\chi}{M_P}\psi \phi_{120}$$
The above example thus illustrates how ansatz such as eq.(\ref{type1deg}) can be realized. A detailed model along this line will require study of the details of the vacuum structure of the potential involving $\eta,\chi$ and possibly additional fields for generating the right structure of the Yukawa  couplings $H,G$.

\section{Numerical Fits: Type-II seesaw}
We now turn to the numerical discussion of the ansatz (\ref{type2deg}) in which the contribution of $\overline{126}$ to fermion masses is assumed to be $O(3)$ invariant. The $O(3)$ breaking arises from the $H$ and $G$ contributions which lead to departure from degeneracy through the type-I seesaw. We shall not specify how this breaking occurs \cite{ka}. Such an ansatz for the type-II contribution was  considered \cite{valle} in  the specific context of $SO(10)$. Detailed fits to fermion masses with recent data are however not presented in these works. We do this essentially following the same procedure as the one adopted in purely type-I case.
We assume $H,G$ to have the most general form. One could choose to work in a basis with a diagonal $H$. In this basis, eq.(\ref{type2deg}) gets changed to
\be
F=c_0 V~,
\ee
where $V$ is a unitary symmetric matrix defined in eq.(\ref{param}). In this basis, the charged fermion mass matrices can be obtained from eq.(\ref{matrices}) by replacing $H$ with diagonal $D_H$, and $F$ with $c_0 V$. The neutrino mass matrix, eq.(\ref{mnu}) can be written in the same basis as
\be
M_{\nu}= m_0 \left(  V - \epsilon ~M_DV^{*}  M_D  ^T\right) 
\ee
The parameter $\epsilon$ controls the contribution from type-I seesaw which induces splittings in neutrino masses.

We use these equations to fit all the fermion masses and mixing using the previous procedure. Results corresponding to the minimal
case are displayed in Table(\ref{tab:3}). The best fit solution we obtained here corresponds to $\chi^2=6.0$ which is acceptable for 16 data points from statistical point of view and all the observables except $m_b$ and $m_s$  are fitted with less than $1 \sigma$ accuracy. The obtained fit in the type-II case is however not as good as in the case of pure type-I seesaw combined with the  ansatz  (\ref{type1deg}).

Numerical fits lead to $\epsilon\approx 2 \times 10^{-6} {\rm GeV}^{-2}$. Since the scale of $M_D$ is set by the top mass the type-I contribution relative to the  type II is given by $\epsilon m_t^2 \sim 10^{-2}$ and type II contribution dominates as assumed. Now the overall scale of the RH neutrino mass is given by (see eq.(\ref{matrices} and ansatz (\ref{type2deg})) 
$$M_3\approx \frac{1}{m_0\epsilon}\approx 2\times 10^{15} {\rm GeV}\left(\frac{0.2 {\rm eV}}{m_0}\right)$$
is close to the GUT scale unlike the minimal models with type-II dominance but hierarchical neutrinos \cite{problem,fits}.  

\begin{table} [h]
\begin{small}
\begin{math}
\begin{array}{|c|c|c|c|}
\hline
 \text{No.} & \text{Observables} & \text{Fitted value}   & \text{Pull} \\
\hline
 1 & m_d\text{[MeV]} & 0.868041  & -0.395023 \\
 2 & m_s\text{[MeV]}  &12.2829 &\textbf{ -1.40714} \\
 3 & m_b\text{[GeV]}  & 1.25634  &\textbf{ 1.69141 }\\
 4 & m_u\text{[MeV]}  & 0.450489  & 0.0032611 \\
 5 & m_c\text{[GeV]}   & 0.210393  & 0.00324503 \\
 6 & m_t\text{[GeV]}   & 102.325  & 0.883371 \\
 7 & m_e\text{[MeV]}  & 0.358502 & 0.00503107 \\
 8 & m_{\mu}\text{[MeV]} & 75.6709  & -0.0111809 \\
 9 & m_{\tau}\text{[GeV]}  & 1.29217  & -0.0244576 \\
 10 &\frac{\dms}{\Dma} & 0.0302538  & 0.00659421 \\
 11 & \sin  \theta _{12}^q & 0.224154  & -0.0913125 \\
 12 & \sin  \theta _{23}^q & 0.0351436  & 0.033571 \\
 13 & \sin  \theta _{13}^q & 0.00326199  & 0.123983 \\
 14 & \sin ^2 \theta _{12}^l & 0.321168  & 0.0214673 \\
 15 & \sin ^2 \theta _{23}^l & 0.439779  & -0.14255 \\
 16 & \sin ^2 \theta _{13}^l & \textbf{0.0356836}  &- \\
 17 & \delta _{CKM}[^{\circ}]& 49.7146 & -0.429864 \\
 18 & \delta _{PMNS}[^{\circ}]&\textbf{ -25.3338}   &  -\\
\hline
&\chi ^2   &   &6.0\\
\hline
\end{array}
\end{math}
\end{small}
\caption{The best fit solution for fermion masses and  mixing obtained assuming the type-II seesaw dominance. Various observables
and their pulls obtained at the minimum are shown (See text for details).}
\label{tab:3}
\end{table}

\section{ Summary} Obtaining a unified description of vastly different patterns of quark and lepton spectrum is a challenging task. This becomes more so if neutrinos are quasi degenerate. We have shown here that it is indeed possible to obtain such a description starting from the fermionic mass structure, eq.(\ref{matrices}) that can arise in a general $SO(10)$ model. We considered two distinct possibilities based on purely type-I and the other based on the mixture of type-I and type-II seesaw mechanisms. Both these possibilities can lead to quasi degenerate spectrum if they are supplemented respectively with ansatz (\ref{type1deg}) and (\ref{type2deg}). We have shown through the detailed numerical analysis that these ansatz are capable of explaining the entire fermionic spectrum and not just the quasi degenerate neutrinos. Moreover, the origin of large leptonic mixing here is linked to the quasi degenerate structure determined by the matrix $V$ providing yet another reason why quark and leptonic mixing angles are so different in spite of underlying unified mass structure.

\end{document}